\documentclass[%
 reprint,
superscriptaddress,
 amsmath,amssymb,
 aps,
pra,
]{revtex4-1}

\usepackage{graphicx}
\usepackage{dcolumn}
\usepackage{bm}
\usepackage{physics}
\usepackage{amsmath}

\newcommand{\bonnpi}{Physikalisches Institut, University of Bonn, Nussallee 12, 53115 Bonn, Germany}
\newcommand{\innsbruck}{Institut f\"ur Theoretische Physik, Universit\"at Innsbruck, Technikerstrasse 21a, A-6020 Innsbruck, Austria}
\begin{document}

\title{Numerically exact treatment of many body self-organization in a cavity}

\date{\today}

\begin{abstract}
   We investigate the full quantum evolution of ultracold interacting bosonic atoms on a chain and coupled  to an optical cavity. Extending the time-dependent matrix product state techniques and the many-body adiabatic elimination technique to capture the global coupling to the cavity mode and the open nature of the cavity, we examine the long time behavior of the system beyond the mean-field elimination of the cavity field.
We investigate the many body steady states and the self-organization transition for a wide range of parameters. We show that in the self-organized phase the steady state consists in a mixture of the mean-field predicted density wave states and excited states with additional defects. In particular, for large dissipation strengths a steady state with a fully mixed atomic sector is obtained crucially different from the predicted mean-field state. 

\end{abstract}
\author{Catalin-Mihai Halati}
\affiliation{\bonnpi}
\author{Ameneh Sheikhan}
\affiliation{\bonnpi}
\author{Helmut Ritsch}
\affiliation{\innsbruck}
\author{Corinna Kollath}
\affiliation{\bonnpi}
\maketitle

Experimental progress to achieve strong coupling of quantum matter to quantum light has opened exciting possibilities. Realizations of such systems nowadays exist both with ultracold atomic gases strongly coupled to optical cavities \cite{ BaumannEsslinger2010, KlinderHemmerich2015,RitschEsslinger2013} or the electron gas in solids coupled to THz cavities \cite{ScalariFaist2012, LiuMenon2015, ZhangKono2016}.
These realizations have allowed one to study self-organization phenomena and stabilize exotic phases by the interaction with the quantum light \cite{RitschEsslinger2013, SmolkaImamoglu2014, BayerLange2017}. 
The advantages of the coupling of quantum matter to quantum light are the fast self-organization dynamics due to the presence of the cavity induced long range interactions and the stabilization of complex states via a dissipative attractor dynamics. 
The cavity induced long-range interaction has been observed  in atomic gases with external optical lattice potentials, where an extended Bose-Hubbard model  has been experimentally realized \cite{KlinderHemmerichPRL2015, LandigEsslinger2016, HrubyEsslinger2018} and the effect of the long-range interactions on the superfluid to insulator transition \cite{MaschlerRitsch2005, MaschlerRitsch2008, LarsonLewenstein2008, NiedenzuRitsch2010, SilverSimons2010, VidalMorigi2010, LiHofstetter2013, ElliotMekhov2016, BakhtiariThorwart2015, FlotatBatrouni2017, LinLode2018} and the out-of-equilibrium dynamics \cite{ChiacchioNunnenkamp2018} have been analyzed. 

Theoretical proposals use the attractor dynamics to stabilize complex quantum phases \cite{MivehvarPiazza2019, OstermannMivehar2019, KiffnerJaksch2019, SchlawinJaksch2019}, including topologically non-trivial phases 
\cite{KollathBrennecke2016, BrenneckeKollath2016, WolffKollath2016, SheikhanKollath2016, ZhengCooper2016, BallantineKeeling2017, HalatiKollath2017, MivehvarPiazza2017}.
Together with the recent achievements regarding the coupling of the cavity field to the internal spin degrees of freedom of atoms \cite{LandiniEsslinger2018,KroezeLev2018, MivehvarRitsch2017}, it has opened the possibilities of the realization of dissipation-induced instabilities  \cite{DograEsslinger2019, ChiacchioNunnenkamp2019, BucaJaksch2019} and dynamical spin-orbit coupling \cite{KroezeLev2019, HalatiKollath2019}.

Previous theoretical descriptions of coupled atomic cavity systems were to a large extent performed using a mean field decoupling of the cavity field and the atoms \cite{RitschEsslinger2013, MaschlerRitsch2008, NagyDomokos2008}, recent efforts have been made to go beyond the mean field description \cite{GammelmarkMolmer2012, WallRey2016, DamanetKeeling2019, SchulerRabl2020}. The mean field approach assumes the cavity field to be in a coherent state and the atoms to be in the ground state of an effective model and can therefore not take the atom-photon coupling correctly into account.  Above a certain threshold of the atom-cavity coupling strength, the cavity field takes a finite value and the atoms self-organize into a non-trivial state. 

So far the exact coupling between the atomic and photonic states has been included only for small systems of one or two atoms, or two sites \cite{VukicsRitsch2007, MaschlerDomokos2007,ZhangZhou2008, KramerRitsch2014, SandnerRitsch2015, OstermannRitsch2019}, non-interacting two-level atoms \cite{XuHolland2013, KirtonKeeling2018, ShammahNori2018}, or in closed systems \cite{PiazzaZwerger2013}. 
In this work, we go beyond the mean field approximation and investigate the combined atom-cavity system developing a quasi-exact numerical simulation based on matrix product states (MPS) and a many body adiabatic elimination approach valid for large photon losses. These methods enable us to study the many body aspects of the self-ordering processes of the interacting bosonic atoms in the optical cavity. 
The dissipative attractor dynamics couples the atoms with the quantum light, even if one starts with a decoupled state of atoms and photons. 
We investigate the nature of the arising steady states for a wide range of parameters. We find that the admixture of excited states beyond the mean field steady state plays an important role in a wide range of parameters. In particular, in the limit of very lossy cavity mirrors the atomic sector approaches the totally mixed state. Our findings question the nature of the pure steady states and phase transitions previously predicted by the zero-temperature mean field theories. The admixtures of excited states in the steady states demonstrates a mixed state nature of the transition and of the steady states.

We consider interacting bosons confined to a chain coupled to a single cavity mode and transversely pumped with a standing-wave laser beam. 
The Lindblad equation for the density operator $\rho$ is given by \cite{Carmichael1993,BreuerPetruccione2002,RitschEsslinger2013, MaschlerRitsch2008}
$\pdv{t} \rho = -\frac{i}{\hbar} \left[ H, \rho \right] + \frac{\Gamma}{2}\left(2a\rho a^\dagger-a^\dagger a \rho-\rho a^\dagger a\right)$,
where $a$ and $a^\dagger$ are the annihilation and creation operators for the photon mode. The term proportional to the dissipation strength $\Gamma$ takes into account the losses from the cavity due to the imperfections of the mirror. 
The first term  represents the unitary evolution in which the excited internal state of the atoms is adiabatically eliminated  \cite{RitschEsslinger2013, MaschlerRitsch2008, NagyDomokos2008}, with $H=H_c+H_{\text{int}}+H_{\text{kin}}+H_{\text{ac}}$, $H_c= \hbar\delta a^\dagger a$, $H_{\text{int}}=\frac{U}{2} \sum_{j=1}^L n_{j}(n_{j}-1)$, $H_{\text{kin}}=-J \sum_{j=1}^{L-1} (b_{j}^\dagger b_{j+1} + b_{j+1}^\dagger b_{j})$, and $H_{\text{ac}}=  -\hbar\Omega ( a + a^\dagger) \Delta, ~~ \Delta=\sum_{j=1}^L (-1)^j n_j $.
The term $H_c$ describes the cavity mode with a detuning between the cavity mode and the transverse pump beam $\delta=\omega_c-\omega_p$, in the rotating frame of the pump beam.
The operators $b_{j}$ and $b_{j}^\dagger$ are the bosonic annihilation and creation operators of the atoms on site $j$ and  $n_{j}=b_{j}^\dagger b_{j}$. $L$ denotes the number of sites of the chain and the total number of atoms is $N$.
$J$ is the tunneling amplitude and $U>0$ the repulsive on-site interaction of strength. 
We assumed a commensurability of the cavity mode with twice the periodicity of the lattice spacing within the chain. This causes the atoms to see different cavity field amplitudes at even and odd sides. As shown in Ref.~\cite{MaschlerRitsch2008}, this leads to a coupling of the cavity field to the total imbalance between the odd and even sites of the chain, $\Delta$, with the effective pump amplitude $\Omega$. In the following we use the scaled coupling strength $\Omega\sqrt{N}$, in order to make our results independent of the particle number.
Whereas typically already to determine the time-evolution of the Bose-Hubbard model alone is very involved, here an additional complication is due to the large and, in principle, unlimited dimension of the Hilbert space of the photonic mode.

This challenge is typically circumvented by adiabatically eliminating the cavity field and using a mean field decoupling for the atoms and the cavity mode  \cite{RitschEsslinger2013}.
Within this crude approximation one finds, that above a certain threshold $\Omega_{\textrm{MF},c}\sqrt{N}$ the cavity field $\langle a \rangle$ takes a finite value, either $\pm \alpha(\Delta)=\frac{\Omega}{\delta-i \Gamma/2}\Delta_{\text{eff}}$ and the atoms self-organize into a density modulated pattern with even-odd imbalance $\Delta_{\text{eff}}$ \cite{sup}. 

We develop here two approaches taking the exact atom-cavity coupling into account.
Both offer new insights into the self-organization of interacting particles and quantum light.

As the first approach, we develop a variant of the many-body adiabatic elimination technique \cite{RipollCirac2009, PolettiKollath2013, HalatiKollath2020} including the photonic mode. This approach is a perturbative approach around the dissipation free subspace and allows us to get analytical insights into the nature of the steady state in the limit of large dissipation , i.e.~ $\hbar\Gamma\gg \hbar\Omega,\,\hbar\delta \gg J$ (see Ref.~\cite{HalatiKollath2020} for details). In particular, the thermodynamic limit can be investigated with this approach. For finite interaction the steady state is given by
$\rho_{\textrm{mix}}=\frac{1}{\mathcal{N}}\sum_{\{n_i\}}\ket{\alpha(\Delta),n_1,\dots,n_L}\bra{\alpha(\Delta),n_1,\dots,n_L}$
\cite{HalatiKollath2020}. The sum runs over all possible density configurations $\{n_i\}$ with $\mathcal{N}$ the total number of atomic configurations, and the coherent state is set by $\alpha(\Delta)=\frac{\Omega}{\delta-i \Gamma/2}\Delta$, where $\Delta$ is taken in $\{n_i\}$. This state, $\rho_\textrm{mix}$, is very distinct from the mean field state and is fully mixed in the atomic sector. 

\begin{figure}[!hbtp]
\centering
\includegraphics[width=.42\textwidth]{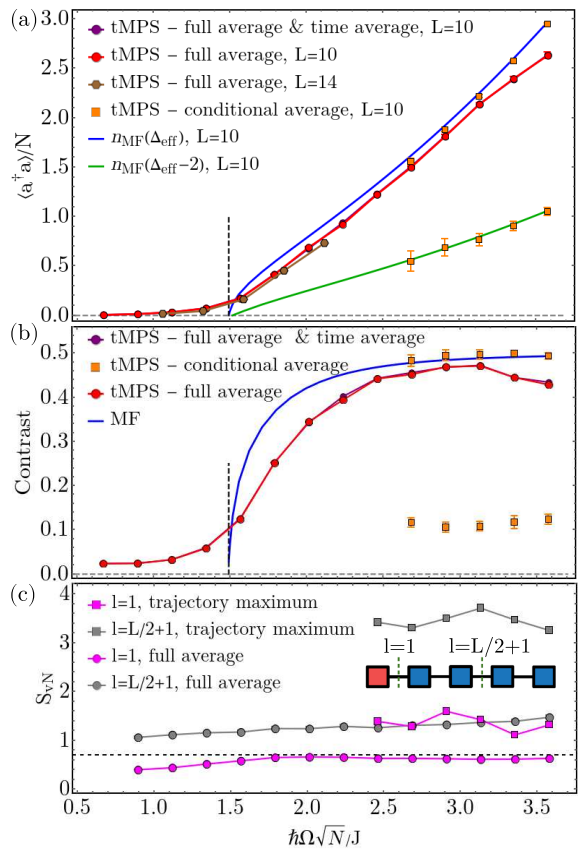}
\caption{
  (a) The scaled photon number, $\langle a^\dagger a\rangle/N$, as a function of $\hbar\Omega\sqrt{N}/J$, for $N/L=1/2$, $\hbar\delta/J=2$, $U/J=2$, and $\hbar\Gamma/J=1$. 
  The purple symbols (below the red symbols) represent a time average for $tJ\in(44.75\hbar,49.75\hbar)$. 
For the orange symbols the trajectories are averaged depending on the final photon number. The blue (green) curves represent the mean-field value of the photon number for the imbalance  $\Delta_{\text{eff}}$ $(\Delta_{\text{eff}}-2)$. The vertical dashed line marks $\Omega_{\textrm{MF},c}\sqrt{N}$. Lines joining the symbols and are guide to the eyes. 
(b) The averaged contrast of the density-density correlation as a function of $\hbar\Omega\sqrt{N}/J$. 
(c) The von Neumann entropy, $S_{\text{vN}}$, as a function of $\hbar\Omega\sqrt{N}/J$, for two bipartitions of the system, between the cavity site and atomic chain (bond $l=1$) and in the middle of the atomic chain (bond $l=L/2+1$). The circles represent the average over all trajectories and the squares the maximum among the trajectories, for $L=10$.
The dashed line represents $\log(2)$.}
\label{fig:photon_no}
\end{figure}

We will show using the second approach how the nature of the long time behaviour changes drastically with the two extreme limits being a state close in nature to either the mean field state, but with a certain admixture of excited states, or to the totally mixed state $\rho_\textrm{mix}$. 
The second approach (for details see Ref.~\cite{HalatiKollath2020}) is a numerically exact treatment of the dissipative time-evolution overcoming the challenges of the long-range and dissipative nature of the photon mode and the presence of interactions within the atoms. This approach combines the Monte-Carlo wave function method \cite{DalibardMolmer1992, GardinerZoller1992} with the matrix product states (MPS)  \cite{ Daley2014,  BernierKollath2013, BonnesLauchli2014}. 
We overcome the challenge of the globally coupled photon mode with a variant which separates off the parts in which the photonic mode occurs by a Trotter-Suzuki decomposition \cite{WhiteFeiguin2004,DaleyVidal2004, Schollwock2011},
 and a dynamical deformation of the MPS structure using swap gates \cite{StoudenmireWhite2010, Schollwock2011, WallRey2016}. A similar variant of the MPS had been applied in the context of spin-boson models \cite{WallRey2016,WallRey2017}, which have no interaction between the spins. 
Our implementation uses the ITensor library \cite{itensor} taking good quantum numbers into account \cite{HalatiKollath2020}. If not stated otherwise, the results are taken at long times $tJ=49.75\hbar$, in order to represent the steady state values \cite{HalatiKollath2020}. The convergence of our results is sufficient  \cite{HalatiKollath2020} for at least 500 trajectories, the truncation error goal of $10^{-12}$, the time-step of $\mathrm{d}tJ/\hbar = 0.0125$ or smaller, a cut-off of the local Hilbert space of the photon mode between $N_\text{pho}=55$ and $N_\text{pho}=10$. The errors bars in all figures represent the standard deviation of the Monte Carlo average. 

\begin{figure}[hbtp]
\centering
\includegraphics[width=.42\textwidth]{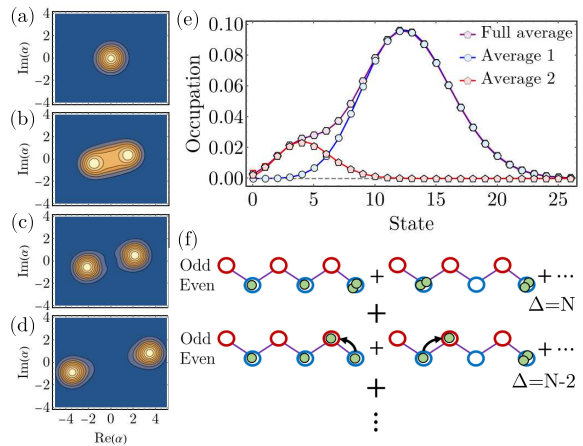}
\caption{(a)-(d) The Q-function for $\hbar\Omega\sqrt{N}/J\in\{1.12,1.79,2.24,3.35\}$, $L=10$, $N=5$, $\hbar\delta/J=2$, $U/J=2$, $\hbar\Gamma/J=1$. 
  (e) Photon number distributions, $p_n=\text{tr}\left(\bra{n}\rho\ket{n}\right)$ for $\hbar\Omega\sqrt{N}/J=3.35$, full average and with the trajectories averaged separately depending on the final photon number. The continuous lines show the Poisson distributions with the corresponding average photon number.
(f) Sketch of the atomic sector of states with perfect imbalance, $\Delta=N$, and states with a reduced imbalance due to a defect, $\Delta=N-2$.
}
\label{fig:2branches}
\end{figure}

We start by analyzing the behavior of the photon number \cite{note1} [Fig.~\ref{fig:photon_no}(a)] in a regime favourable for the mean field treatment. 
A smooth increase in the photon number across the self-organization threshold predicted by mean field is seen which does not show strong system size dependence. However, above the threshold the values of our numerical results remain below $n_{\textrm{MF}}$.
We will show later that this has its origin in the admixture of states with a reduced photon number. In order to get more insight into the obtained state, we study the phase space distribution of the cavity field, represented by the Q-function, $Q(\alpha)=\text{tr}\left(\bra{\alpha}\rho\ket{\alpha}\right)$. 
We can observe in Fig.~\ref{fig:2branches}(a), that for $\hbar\Omega\sqrt{N}=1.12 J$ $Q(\alpha)$ has a maximum at $\alpha=0$ which resembles a coherent state with zero photons. In contrast, above the threshold the Q-function develops two maxima [Fig.~\ref{fig:2branches}(b)] which separate
(Fig.~\ref{fig:2branches}(c)).
At large $\Omega\sqrt{N}$ [Fig.~\ref{fig:2branches}(d)], both peaks in $Q(\alpha)$ deviate from the circular shape and states with a lower photon number are populated.

The atomic part of the steady state above the mean field threshold, shows the characteristic staggered density-wave in the density-density correlations.
In Fig.~\ref{fig:photon_no}(b) we quantify this staggering, by the average contrast by $\frac{1}{L-2}\sum_j\left(\langle n_{j}n_{j+2}\rangle-\langle n_{j}n_{j+1}\rangle\right)$. Across the self-organization threshold the contrast shows a strong increase indicating the build up of a  density wave. However, above the threshold our numerical results remain below the mean field prediction. 

The von Neumann entropy  $S_{\text{vN}}$  in the quantum trajectories  [Fig.~\ref{fig:photon_no}(c)] \cite{note2, HalatiKollath2020} measures the entanglement present in each trajectory. Since the presence of entanglement typically limits the possible compression in the MPS methods, the von Neumann entropy is one of the crucial convergence parameters of these methods \cite{Schollwock2011}. We find that $S_{\text{vN}}$ is finite and saturates in time. Thus, we can be confident that our approach captures the dynamics of the system correctly. In Fig.~\ref{fig:photon_no}(c) we see $S_{\text{vN}}$ computed between the photon mode and the atomic chain seems to be independent of $\Omega$ above the threshold and close to $\log(2)$. We attribute this value to the coherent superposition of the two states corresponding to a different sign of the photon field in each trajectory. 

We analyze the origin of the deviations from mean field by considering the single quantum trajectories. The trajectories stabilize at two different photon numbers. Thus, we implemented a conditional averaging process, depending on the final photon number. The two obtained photon number distributions [Fig.~\ref{fig:2branches}(e)], agree approximately with a Poisson distribution with the corresponding average photon number. 
In contrast to the full average, the expectation value averaged over the first class of trajectories of the photon number [Fig.~\ref{fig:photon_no}(a)] and the staggering contrast [Fig.~\ref{fig:photon_no}(b)] agree well with the mean field prediction.
Thus, the state resembles a good charge density wave in the first class of trajectories. 

In contrast, we attribute the second class of trajectories to states which have an additional defect due to the tunneling of an atom. In the limit of perfect imbalance $\Delta_{\text{eff}}=N$, these states would have only one atom at the "wrong" site [Fig.~\ref{fig:2branches}(f)].
More generally, the reduced average value of the photon number can  [Fig.~\ref{fig:photon_no}(a)] be well explained assuming that the imbalance is reduced as $\Delta\approx\Delta_{\text{eff}}-2$. 
The photon number distribution resembles a coherent state with this lower photon number [Fig.~\ref{fig:2branches}(e)].
We can distinguish between the two types of trajectories only for $\hbar\Omega\sqrt{N}/J\geq 2.68$, as for lower pump strengths the individual quantum trajectories are too noisy due to the low photon number. The presence of the trajectories belonging to two states different in nature strongly suggests that the numerically observed steady state is a mixture of these two dominant contributions. This is further confirmed by exact diagonalization studies on small systems which show a unique steady state being the mixture of the identified states. 
Therefore, a crucial deviation from the mean field predictions of a pure state transition is identified.

\begin{figure}[hbtp]
\centering
\includegraphics[width=.42\textwidth]{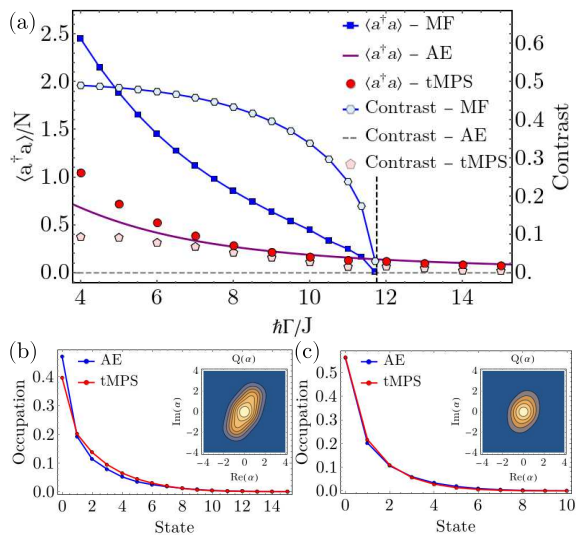}
\caption{ (a) The scaled photon number, $\langle a^\dagger a\rangle/N$ and the averaged contrast of the density-density correlation, $\frac{1}{L-2}\sum_j\left(\langle n_{j}n_{j+2}\rangle-\langle n_{j}n_{j+1}\rangle\right)$, as a function of $\hbar\Gamma/J$ using tMPS, mean-field (MF) and many-body adiabatic elimination (AE).
(b)-(c) The full photon number distribution, $p_n=\text{tr}\left(\bra{n}\rho\ket{n}\right)$ for (b) $\hbar\Gamma/J=7.5$ and (c) $\hbar\Gamma/J=10$. The insets present the corresponding Q-function determined by tMPS. The parameters are chosen to be $L=10$, $N=5$, $\hbar\Omega\sqrt{N}/J=4.47$, $\hbar\delta/J=2$, $U/J=2$.
}
\label{fig:ae}
\end{figure}

The deviations from the mean field predictions become even more prominent in the regime of strong dissipation. We attribute this to the admixture of states which correspond to more and more defects until in the limit of very large dissipation $\Gamma$ the state $\rho_{\text{mix}}$  is reached. We can observe that for a large $\Gamma$ the photon number does no longer agree with the mean field value, but matches fairly well with the value computed for $\rho_{\text{mix}}$ [Fig.~\ref{fig:ae}(a)]. In particular, whereas the mean field approach predicts that at $\hbar\Gamma/J\approx 11.6$  a transition back to the normal phase occurs, we do not observe this transition, as the photon number remains finite in the numerical results \cite{HalatiKollath2020}. The agreement with the adiabatic elimination results becomes very good also in the distribution of the photon number [Fig.~\ref{fig:ae}(b)-(c)]. Where at $\hbar\Gamma/J=7.5$ still small deviations are present at low number states, the distribution for $\hbar\Gamma/J=10$ agrees almost perfectly. The Q-function no longer has two maxima at large $\Gamma$ (insets of Figs.~\ref{fig:ae}(b)-(c)), but only one maximum at $\alpha=0$ and a squeezed profile.

The same agreement of our numerical results and the adiabatic elimination state can be seen in the staggered contrast of the density-density correlations. For $\rho_{\text{mix}}$ the contrast in the staggering vanishes. Increasing $\Gamma$, we see that the contrast approaches zero [Fig.~\ref{fig:ae}(a)]. Thus, at large values of the photon losses, the self-organized steady state no longer resembles a staggered density wave state. It is a mixture with a contribution from many atomic and photonic states, but where each atomic state fully determines the state of the cavity field.

In the thermodynamic limit the adiabatic elimination state, $\rho_\text{mix}$, predicts that the scaled average photon number $\langle a^\dagger a\rangle/N$ goes to zero \cite{HalatiKollath2020}. This would correspond to the mean field predictions of having a transition back to an empty cavity. However, even though the average value of the scaled photon number vanishes, for the adiabatic elimination state this is associated to the admixture of more and more defect states. In the atomic sector the state corresponds to a fully mixed infinite temperature state, as already seen in the reduced average contrast of the density-density correlations. Therefore also in the thermodynamic limit the obtained state is very different from the expected pure mean field state. Our findings rise the question whether a phase transition is expected in the thermodynamic limit. In particular, if such a transition exists, our results suggest that the nature of this transition would not correspond to a zero-temperature phase transition, but that the transition would be dominated by the admixture of excited states. 

In summary, we performed the full quantum time-evolution towards the many body steady state of a chain of interacting bosonic atoms coupled to an optical cavity. We showed that by including the coupling between the atomic degrees of freedom and the photonic field one finds important deviations from the mean field approach of eliminating the cavity field. We saw that when the dissipation strength is comparable with the other energy scales in our system, the system is in a mixture where the largest contribution is given by a density wave state. Other states without density ordering become more prominent in the mixture as we increase the dissipation strength, such that in the large $\Gamma$ limit the atomic sector is fully mixed, but with a strong coupling between the atomic and the photonic sector. This questions the previous picture obtained by zero-temperature mean field theories which assume pure state transitions and replaces it by transitions of a mixed state character. 

We verified that in the experimental parameter regimes of the current realizations Refs.~\cite{KlinderHemmerichPRL2015, LandigEsslinger2016, HrubyEsslinger2018, KroezeLev2018} the predicted mixed character of the transition and steady states occurs in the considered one-dimensional systems. A first sign of these mixed state transitions would be the finding of single experimental runs which stabilize at different photon numbers. However, in order to uniquely detect the mixed state character in the atomic sample, an additional observable as the direct measurement of the even-odd-site imbalance and density-density correlations of the atoms would be desirable. This can by now be measured in optical lattice setups in the absence of a cavity and we expect that our findings motivate the experiments to implement this in the cavity setups.

\textit{Acknowledgments}: We thank J.-S. Bernier, T. Donner,  M. K{\"o}hl, S. Ostermann, F. Piazza, U.~Schollw\"ock, S.~Wolff, W.~Zwerger
for stimulating discussions. We acknowledge funding from the Deutsche Forschungsgemeinschaft (DFG, German Research Foundation) in particular under project number 277625399 - TRR 185 (B4), project number 277146847 - CRC 1238 (C05), FOR1807 and under Germany's Excellence Strategy – Cluster of Excellence Matter and Light for Quantum Computing (ML4Q) EXC 2004/1 – 390534769 and the European Research Council (ERC) under the Horizon 2020 research and innovation programme, grant agreement No.~648166 (Phonton).


\begin{thebibliography}{10}
\bibitem{note1}
The mean-field order parameter vanishes always in our numerical results at long times, i.e.~ $\langle a \rangle = 0$.
\bibitem{note2}
We note that the von Neumann entropy averaged over the quantum trajectories is not a measure of the entanglement present in the resulting density matrix.
\end{thebibliography}
\end{document}